\documentclass{osa-article}

\journal{osac}

\usepackage{csquotes}
\usepackage{amsmath}
\usepackage{subfig}

\articletype{Research Article}
\usepackage{verbatim}
\begin{document}

\title{Entanglement of microwave and optical fields using electrical capacitor loaded with plasmonic graphene waveguide}

\author{Montasir Qasymeh\authormark{1,*} and Hichem Eleuch,\authormark{2,3}}

\address{\authormark{1}Electrical and Computer Engineering Department, Abu Dhabi University, Abu Dhabi, UAE\\
\authormark{2} Department of Applied Sciences and Mathematics, College of Arts and Sciences,\\ Abu Dhabi University, UAE\\
\authormark{3}Institute for Quantum Science and Engineering, Texas AM University, Texas, USA}

\email{\authormark{*}montasir.qasymeh@adu.ac.ae} 


\begin{abstract}
We propose a novel approach for microwave and optical fields entanglement using an electrical capacitor loaded with graphene plasmonic waveguide. In the proposed scheme, a quantum microwave signal of frequency $\omega_{m}$ drives the electrical capacitor, while an intensive optical field (optical pump) of frequency $\omega_1$ is launched to the graphene waveguide as surface plasmon polariton (i.e., SPP) mode. The two fields interact by the means of electrically modulating the graphene optical conductivity. It then follows that an upper and lower SPP sideband modes (of $\omega_{2}=\omega_{1}+\omega_{m}$ and $\omega_{3}=\omega_{1}-\omega_{m}$ frequencies, respectively) are generated. We have shown that the microwave signal and the lower sideband SPP mode are entangled, given a proper optical pump intensity is provided. A quantum mechanics model is developed to describe the fields evolution. The entanglement of the two fields is evaluated versus many parameters including the waveguide length, the pump intensity, and the microwave frequency. We found that the two fields are entangled over a vast microwave frequency range. Furthermore, our calculations show that a significant number of entangled photons are generated at the lower SPP sideband. The proposed scheme attains tunable mechanism for microwave-optical entanglement which paves the way for efficient quantum systems.
\end{abstract}

\section{Introduction}
Entanglement $\big($which is a quantum correlation that exceeds classical limits regardless the separating distance \cite{violation}, \cite{BellInequality} $\big)$ has enabled many unprecedented applications. These include (but not limited to) quantum teleportation \cite{Teleportation1},\cite{Teleportation2}, satellite quantum communication \cite{Satellite}, submarine quantum communication \cite{submarine}, quantum internet \cite{internet}, quantum error correction \cite{correction}, quantum cryptography \cite{Cryptography}, just to mention few.

Several reports have investigated entanglement in different configurations. These include two optical fields entanglement using beam splitter \cite{splitter}, \cite{splitterHichem} (or nonlinear medium \cite{Nonlinear}, \cite{NonlinearHichem}), two trapped ions entanglement \cite{Ions}, entanglement of optical photon and phonon pair \cite{phonon}, entanglement of two optomechanical systems \cite{2optomechanical}, \cite{2optomechanicalHichem} optical photon entanglement with electron spin \cite{spin}, entanglement of mechanical motion with microwave field \cite{motion}, entanglement of micormechanical resonator with optical field \cite{micromechanical}, \cite{micromechanicalHichem}, and entanglement of two microwave radiations \cite{radiation},\cite{radiation2}. Furthermore, recent reports have proposed schemes for microwave and optical fields entanglement \cite{MicroOptic1}-\cite{MicroOptic3}. As a mater of fact, achieving entangled microwave and optical fields is very vital to combined superconductivity with quantum photonic systems \cite{combined}, which enables efficient quantum computation and communications. In \cite{MicroOptic1}, the entanglement between microwave and optical fields were achieved by means of mechanical resonator coupling between the two fields. While using quantum mechanical resonator limits the frequency tunability, the major drawback of this approach is the sensitivity of the mechanical resonator to thermal noise. A different approach is presented in \cite{MicroOptic2}. The entanglement between microwave and optical fields is conducted using an optoelectronic system (compressed of a photodetector and a Varactor diode). While this approach avoids the thermal noise restriction and can be designed to be tunable, the bandwidth of the photodetector and the Varactor capacitor (and their noise figures) imposes the performance limitations. A recent approach is proposed in \cite{MicroOptic3} for microwave and optical fields entanglement using whispering gallery mode resonator filled with electro-optical material. In this approach, an optical field is coupled to the whispering gallery resonator while a microwave field drives the resonator. There are several constraints that must be met, though. First, the driving microwave field and the optical mode in the whispering gallery resonator must be well overlapping to conduct the interaction. Also, a sophisticated coupling approach is needed to launch the optical field into the whispering resonator. It then follows that the operation must be optimized for specific microwave and optical frequencies. Second, the free spectral range of the whispering resonator must match the microwave frequency, which also limits tunability. Third, the size of the whispering resonator needs to be in the millimeter range (i.e., bulky) to attain high quality factor. Thus, in the light of the above, a novel approach (with an off-resonance mechanism) is needed to achieve a wide band entanglement of microwave and optical fields with large tunability.  

In this work, we proposed a novel approach for microwave and optical fields entanglement based on electrical capacitor loaded with graphene plasmonic waveguide. As the microwave signal drives the parallel plates of the capacitor, the garphene waveguide supports a surface plasmon polariton (i.e., SPP) mode. The microwave voltage and the SPP mode interact by the means of electrically modifying the graphene optical conductivity. In this work, we consider an optical SPP pump of frequency $\omega_{1}$ and a microwave signal of frequency $\omega_{m}$. It then follows that an optical SPP sidebands at frequencies $\omega_{2}=\omega_{1}+\omega_{m}$ and $\omega_{3}=\omega_{1}-\omega_{m}$ are generated. We show that the driving microwave signal and the lower sideband at $\omega_{3}$ are entangled for proper pump intensity $\lvert A_1 \rvert ^2$. We have evaluated the entanglement of the microwave and the optical field versus different parameters including the graphene waveguide length, the microwave frequency, the microwave number of photons and the pump intensity. We found that entanglement is achieved (and can be tuned) over vast microwave frequency range given proper pump intensity is supported.   

The rest of the paper is organized as in the following: In section 2, the description of the proposed structure (and the pertinent propagating SPP modes) are presented. In section 3, a quantum mechanics model is developed. Section 4 discusses the entanglement between the microwave and the SPP lower sideband. The numerical evaluations are presented in section 5. Section 6, addresses the conclusion remarks.   

\begin{figure}[ht!]
\centering\includegraphics{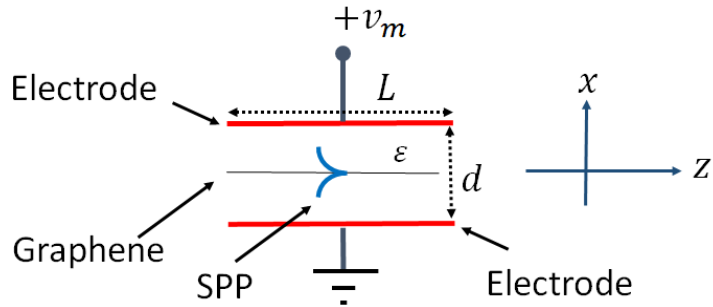}
\caption{The proposed structure of electrical capacitor loaded with plasmonic graphene waveguide}
\end{figure}

\section{Proposed Structure}
Consider a superconducting parallel plate capacitor loaded with graphene layer, as shown in Fig. 1. The two plates are separated by distance $d$, lie in the $yz$ plane, and have $\mathcal{A}_r=L\times W$ area. The graphene layer is located in the middle between the two plates at $z=0$. The capacitance (per unit area) is given by $C=\frac{\varepsilon \varepsilon_0}{d}$.  

The capacitor is driven by a quantum microwave signal, that is:
\begin{equation} \label{eq1}
V_m=\mathcal{V} e^{-i\omega_mt}+c.c.
\end{equation}
A transverse magnetic (i.e., TM) surface plasmon polariton (i.e., SPP) mode is coupled to the graphene waveguide. The SPP mode is described by its associated electrical (and magnetic) fields, given by:

\begin{equation}\label{eq2}
\Vec{E}=\mathcal{U}(z) \Big(\mathcal{D}_x(x) \Vec{e_x}+\mathcal{D}_z(x) \Vec{e_z}\Big) e^{-i\big(\omega t-\beta z\big)}+c.c., 
\end{equation}

\begin{equation}
\Vec{H}=\mathcal{U}(z) \mathcal{D}_y(x) \Vec{e_y} e^{-i\big(\omega t-\beta z\big)}+c.c.,
\end{equation}
where, $\mathcal{U}(z) $ is the complex amplitude, $\mathcal{D}_x(x)=\Big\{ \frac{\beta i}{\omega \varepsilon \varepsilon_0} e^{\alpha x}$ for $x< 0;\; \frac{\beta i}{\omega \varepsilon \varepsilon_0} e^{-\alpha x}$ for $x>0 \Big\}$, \; $\mathcal{D}_z(x)=\Big\{\frac{\alpha i}{\omega \varepsilon \varepsilon_0} e^{\alpha x}$ for $x< 0; \; \frac{\alpha i}{\omega \varepsilon \varepsilon_0} e^{-\alpha x}$ for $x>0 \Big\}$ and $\mathcal{D}_y(x)=\Big\{ e^{\alpha x}$ for $x< 0; \; e^{-\alpha x}$ for $x>0 \Big\}$ are the spatial distributions of the SPP mode,  $\alpha =\sqrt{\beta^2-\varepsilon k_0^2}$, and $k_0=\frac{\omega}{c}$ is the free space propagation constant and $c$ is the speed of light in the vacuum.

The dispersion relation of the SPP mode is given by:  
\begin{equation} \label{eq3}
\beta=k_0 \sqrt{1-\big(\frac{2}{Z_0\sigma_s}\big)^2},
\end{equation}
where $Z_0=377 \Omega$ is the free space impedance, and $\sigma_s$ is the graphene conductivity (see Appendix A).

For an input SPP mode of frequency $\omega_1$ and a driving microwave voltage of frequency $\omega_m$, an upper and lower  SPP sidebands are generated at frequencies $\omega_2=\omega_1+\omega_m$ and $\omega_3=\omega_1-\omega_m$ by the means of graphene conductivity modulation \cite{qasymehhichem}, \cite{qasymehphase}. The associated electric fields with these SPP modes are given by $\Vec{E}_{j}=\mathcal{U}_j(z) \Big( \mathcal{D}_{xj}(x) \Vec{e_{x}}+\mathcal{D}_{zj}(x) \Vec{e_z}\Big) e^{-i\big(\omega_j t-\beta_j z\big)}+c.c.$, where $j \in \{1,2,3\}$. 
On implementing a perturbation approach (see details in Appendix A), the effective propagation constant of the SPP modes can be approximated by $\beta_j=\beta_{j}^{\prime}+ \mathcal{V} \beta_{j}^{\prime\prime} e^{-i \omega_m t}+c.c$, and thus, the corresponding effective permittivity of the SPP modes is given by: 

\begin{equation} \label{eq4}
\varepsilon_{eff_j}=\varepsilon_{eff_j}^{\prime}+ \mathcal{V}\varepsilon_{eff_j}^{\prime\prime} e^{-i \omega_m t}+c.c.,
\end{equation}
where $\varepsilon_{eff_j}^{\prime}=\bigg(\frac{\beta_j^{\prime}}{k_{0_j}}\bigg)^2$, $ \varepsilon_{eff_j}^{\prime\prime}=2\frac{\beta_j^{\prime} \beta_j^{\prime\prime}}{k_{0_j}^2}$, $\beta_j^{\prime}$ is the solution of the dispersion relation in Eq.(\ref{eq3}), $ \beta_j^{\prime\prime}=\frac{\beta_j^{\prime}}{1-\Big(\frac{1}{2} Z_0 \sigma_{s_j}^{\prime}\Big)^2}\frac{\sigma_{s_j}^{\prime\prime}}{\sigma_{s_j}^{\prime}}$, and $\sigma_{s_j}^{\prime\prime}$ is the perturbed graphene conductivity term (defined in Appendix A). 

The above presented model implies that the SPP modes are contained between the two plates with negligible overlapping with the electrodes. This can be attained by having the separating distance between the two electrodes $d$ adequately larger than $\frac{1}{\alpha}$. For example, for $d=\frac{10}{\alpha}$, then $99.99\%$ of the SPP mode is contained within the gap between the two parallel plates \cite{qasymehTHz}.    

\section{Quantum Mechanics Description}

The interacting fields can be quantized through the following relations:

\begin{equation} \label{eq7}
	\mathcal{U}_j= \frac{\big(\hbar \omega_j\big)^{\frac{1}{2} }}{ \xi_j^{\frac{1}{2}}\bigg(\varepsilon_{0}\varepsilon_{eff_{j}}^{\prime} V_L \bigg)^{\frac{1}{2}}} \hat{a}_j,  \quad \textrm {and} \quad \mathcal{V}= \bigg( \frac{2\hbar \omega_m}{ C \mathcal{A}_r}\bigg)^{\frac{1}{2}} \hat{b},
\end{equation}
where $\hat{a}_j$ and $\hat{b}$ are the annihilation operators of the $ j^{th}$ optical and  microwave fields, respectively, $ V_L=  \mathcal{A}_r \int_{\mathcal{-\infty}}^{+\infty}\big( \lvert \mathcal{D}_{x_j}\rvert\ ^{2} +\lvert \mathcal{D}_{z_j}\rvert\ ^{2} \big) \partial x $ is the SPP volume, $\xi_j=\frac{1}{2}+ \frac{\mu_0}{2\varepsilon_0 \varepsilon_{eff_j}^\prime} \frac{\int_{\mathcal{-\infty}}^{+\infty} \lvert \mathcal{D}_{y_j} \rvert\ ^2 \partial x}{\int_{\mathcal{-\infty}}^{+\infty} \big( \lvert \mathcal{D}_{x_j} \rvert\ ^2 + \lvert D_{z_j} \rvert\ ^2 \big) \partial x}$ is a unit-less parameter that is introduced to match the expression of the free Hamiltonian of the SPP modes (i.e.,$\hat{\mathcal{H}_0}$ ) to the expression of the free Hamiltonian of the corresponding unguided fields. It then follows that the spatial distribution of the SPP modes is completely included in the conversion rates $g_2$ and $g_3$. 

Consequently, by substituting the relations in Eqs.(\ref{eq7}) into Eq.(\ref{eqB9}), the quantum Hamiltonian is given by:


\begin{equation} \label{eq8}
	\hat{\mathcal{H}} = \hat{\mathcal{H}_0}+\hat{\mathcal{H}_1},
\end{equation}

where 
\begin{equation} \label{eq9}
	\hat{\mathcal{H}_0} =\hbar \omega_m \hat{b}^\dagger \hat{b}+\sum_{j=1}^{3} \hbar \omega_j \hat{a}_{j}^\dagger \hat{a}_j, \quad \textrm {and} \quad \hat{\mathcal{H}_1} =\hbar g_2 \hat{a}_{2}^\dagger \hat{b} \hat{a}_{1} +\hbar g_3 \hat{a}_{1}^\dagger \hat{b} \hat{a}_{3} +h.c.,
\end{equation}

$h.c.$ is the Hermitian conjugate, and $g_2$ and $g_3$ are the conversion rates given by:

\begin{equation} \label{eq11}
g_2 = \frac{1}{2}\varepsilon_{eff_2}^{\prime\prime} sinc \bigg(\frac{\beta_1-\beta_2}{2}L\bigg) e^{i\frac{\beta_1-\beta_2}{2}L}
	\bigg( \frac{2 \omega_1 \omega_2 \hbar \omega_m }{C \mathcal{A}_r  \varepsilon_{eff_1}^{\prime}\varepsilon_{eff_2}^{\prime}}\bigg)^\frac{1}{2} \frac{I_{12}}{\sqrt{\xi_1\xi_2}},
\end{equation}

\begin{equation} \label{eq12}	
	g_3 = \frac{1}{2}\varepsilon_{eff_3}^{\prime\prime} sinc \bigg(\frac{\beta_3-\beta_1}{2}L\bigg) e^{i\frac{\beta_3-\beta_1}{2}L}
	\bigg( \frac{2 \omega_3 \omega_1 \hbar \omega_m }{C \mathcal{A}_r \varepsilon_{eff_1}^{\prime}\varepsilon_{eff_3}^{\prime}}\bigg)^\frac{1}{2}  \frac{I_{13}}{\sqrt{\xi_1\xi_3}}.
\end{equation}
where $I_{mn}=\frac{\int_{\mathcal{-\infty}}^{+\infty} \big(  \mathcal{D}_{x_m}^* \mathcal{D}_{x_n}+\mathcal{D}_{z_m}^* \mathcal{D}_{z_n} \big)  \partial x }{\sqrt{\int_{\mathcal{-\infty}}^{+\infty} \big(\lvert \mathcal{D}_{x_m} \rvert\ ^2 +\lvert \mathcal{D}_{z_m} \rvert\ ^2 \big)\partial x} \sqrt{\int_{\mathcal{-\infty}}^{+\infty} \big(\lvert \mathcal{D}_{x_n} \rvert\ ^2 +\lvert \mathcal{D}_{z_n} \rvert\ ^2 \big)\partial x}}$.

The SPP pump at frequency $\mathcal{\omega}_1$ is intensive and treated classically. It then follows that on substituting the quantum Hamiltonian expression of Eq. (\ref{eq8}) into the Heisenberg equations of motion, that is $\frac{\partial\hat{x}}{\partial t}=\frac{i}{\hbar}  [\hat{\mathcal{H}},\hat{x}]$, and using the rotation approximation (i.e., $\hat{o_j}=\hat{O_j} e^{-i\mathcal{\omega_j} t}$), one yields the following equations of motion: 

\begin{equation} \label{eq13}
   \frac{\partial\hat{A}_2}{\partial t}=-\frac{\Gamma_2}{2} \hat{A}_{2}+ g_2 A \hat{B}+\sqrt{\Gamma_2}\hat{N}_2,
\end{equation}

\begin{equation} \label{eq14}
   \frac{\partial\hat{A}_3}{\partial t}=-\frac{\Gamma_3}{2} \hat{A}_{3}+ g_3 A \hat{B}^{\dagger}+\sqrt{\Gamma_3}\hat{N}_3,
\end{equation}

\begin{equation} \label{eq15}
   \frac{\partial\hat{B}}{\partial t}=-\frac{\Gamma_m}{2} \hat{B}- g_2 A^* \hat{A}_{2}+g_3 A \hat{A}_{3}^{\dagger}+\sqrt{\Gamma_m}\hat{N}_m,
\end{equation}
where $\Gamma_j= 2 v_g Im{(\beta^{\prime})}$ is the optical decay coefficient, $ \Gamma_m$ represents the microwave decay coefficient, and $v_g=\frac{\partial f}{\partial \beta}$ is the group velocity. Here, the pump field amplitude $A_1$ is considered with $\frac{\pi}{2}$ phase (i.e., $ A_1 = A e^{i\frac{\pi}{2}}=iA$) for seek of simplicity, and $N_2$ and $N_m$ are the quantum Langevin noise operators. The dissipation is characterized by the time decay rates, which are included in the equation of motion in Eqs. (\ref{eq13} to \ref{eq15}). Hence, according to the fluctuation-dissipation theorem, the Langevin forces, i.e., $\hat{N}_j$ , are also included. The quantum coupled equation of motion presented above describe the evolution of the SPP modes and the driving microwave signal. In the following sections we investigate the entanglement between the microwave and optical SPP modes. Such a quantum phenomenon would pave the way for novel quantum microwave photonic systems.        


\section{Entangled Microwave and Optical Fields}

As can be seen from the motion equations (Eqs. \ref{eq14} and \ref{eq15}), the microwave annihilation (creation) operator $\big($i.e., $\hat{B}$ ($\hat{B}^{\dagger}$)$\big)$ is coupled to the SPP lower side band creation (annihilation) operator $\big($i.e., $\hat{A}_{3}^{\dagger}$ ($\hat{A}_{3}$)$\big)$, which implies possibility for entanglement. Several techniques have been developed to quantify entanglement. These include logarithmic negativity \cite{negativity}, \cite{negativityHichem}, the degree of Einstein-Podolsky-Rosen (EPR) paradox \cite{EPR}, Peres-Horodecki criterion\cite{Horodecki}, and inseparability Duan's criterion \cite{Duan}, \cite{Duan2}. 

In this work, no steady state can be considered as the interaction is carried out while the propagating SPP modes are coupled to optical pump. Thus, the time rate of the SPP modes averages are nonzero $ \big ( \frac{ \partial \left\langle \hat{A}_{j}\right\rangle}{\partial t}\neq 0 \big )$. To address these requirements, we obey the following approach to evaluate the entanglement between $\hat{B}$ and $\hat{A}_3$. 
First, we consider the Duan's criterion in the determinant form (Eq. \ref{eq16}). It then follows that the entanglement is existing whenever the determinant is negative (i.e., $\Lambda < 0$ ), \cite{Duan}.
\begin{equation} \label{eq16}
\Lambda=
\begin{vmatrix}
1 &\left\langle \hat{A}_{3}\right\rangle& \left\langle \hat{B}^{\dagger}\right\rangle \\ 
\left\langle \hat{A}_{3}^{\dagger}\right\rangle& \left\langle \hat{A}_{3}^{\dagger}\hat{A}_{3}\right\rangle & \left\langle \hat{A}_{3}^{\dagger}\hat{B}^{\dagger}\right\rangle \\
 \left\langle \hat{B}\right\rangle&  \left\langle \hat{A}_{3}\hat{B}\right\rangle   & \left\langle \hat{B}^{\dagger}\hat{B}\right\rangle
\end{vmatrix},
\end{equation}
Second, we obtain the rate equations for the operators' averages $\big($ by applying the average operator to Eqs.(\ref{eq13}-\ref{eq15})$\big)$, yielding:   

\begin{equation} \label{eqC1}
   \frac{\partial\left\langle \hat{A}_{2}\right\rangle}{\partial t}=-\frac{\Gamma_2}{2} \left\langle \hat{A}_{2}\right\rangle+ g_2 A \left\langle \hat{B}\right\rangle,
\end{equation}

\begin{equation} \label{eqC2}
   \frac{\partial\left\langle \hat{A}_{3}\right\rangle}{\partial t}=-\frac{\Gamma_3}{2} \left\langle \hat{A}_{3}\right\rangle + g_3 A \left\langle \hat{B^\dagger}\right\rangle,
\end{equation}

\begin{equation} \label{eqC3}
   \frac{ \partial\left\langle \hat{B}\right\rangle}{\partial t}=-\frac{\Gamma_m}{2} \left\langle \hat{B}\right\rangle- g_2 A^* \left\langle \hat{A}_2\right\rangle +g_3 A \left\langle \hat{A}_3^\dagger \right\rangle,
\end{equation}
Third, we obtain the rate equations for $\left\langle \hat{A}_{3}^{\dagger}\hat{A}_{3}\right\rangle$, $\left\langle \hat{A}_{3}^{\dagger}\hat{B}^{\dagger}\right\rangle$,$\left\langle \hat{A}_{3}\hat{B}\right\rangle$ and $\left\langle \hat{B}^{\dagger}\hat{B}\right\rangle$, using the quantum regression theorem $\big ($ see  Eqs.\ref{eqC4} to \ref{eqC11} in Appendix B $\big )$ \cite{qasymehhichem}.
Fourth, we use numerical iterative approach (i.e., finite difference method) to solve the coupled differential equation set in (Eq.\ref{eqC1} to Eq. \ref{eqC3}) and in (Eq.\ref{eqC4} to Eq.\ref{eqC11}) to obtain the required values to evaluate the condition in Eq.\ref{eq16} at specific interaction time $t=\frac{L}{v_g}$. The microwave and optical operators are consider uncorrelated at time $t=0$, which implies that $\left\langle \hat{A}_j^{\dagger}\hat{B}^{\dagger}\right\rangle |_{t=0}=\sqrt{\left\langle \hat{B}^{\dagger}\hat{B}\right\rangle }|_{t=0}\sqrt{\left\langle \hat{A}_j^{\dagger}\hat{A}_j\right\rangle }|_{t=0}$ and  $\left\langle \hat{A}_j\hat{B}\right\rangle |_{t=0}=\sqrt{\left\langle \hat{B}^{\dagger}\hat{B}\right\rangle }|_{t=0}\sqrt{\left\langle \hat{A}_j^{\dagger}\hat{A}_j\right\rangle }|_{t=0}$. Here, $\left\langle \hat{A}_3^{\dagger}\hat{A}_3\right\rangle |_{t=0}=0$, $\left\langle \hat{A}_2^{\dagger}\hat{A}_2\right\rangle |_{t=0}=0$, and $\left\langle \hat{B}^{\dagger}\hat{B}\right\rangle |_{t=0}$ is the number of microwave photons at $t=0$.
In the following section, the entanglement of the two fields ($\hat{B}$ and $\hat{A}_{3}$) is numerically evaluated versus different parameters, including the waveguide length, the SPP pump intensity, the microwave number of photons, and the microwave frequency.

\begin{figure}[ht!]\label{dispersion}
\centering\includegraphics[width=7cm]{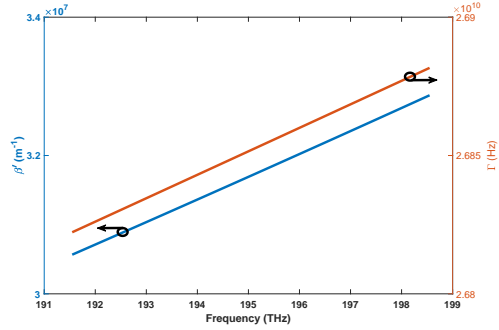}
\caption{The propagation constant and the decay time of the SPP mode versus the optical frequency}
\end{figure}

\section{Results and Discussion}
In this section, we present numerical evaluations of our proposed entanglement scheme considering practical parameters. The electrical capacitor is considered with air filling material. The graphene doping concentration is $n_0=10^{18} m^{-3}$, the pump frequency is $\frac{\omega_1}{2\pi}=$193 THz, and the temperature is $T=3 mK$. 

Using these parameters, the SPP propagation constant $\beta$ (and the decay time constant $\Gamma$) are presented, in Fig. 2, versus the optical frequency. Consequently, by calculating $\alpha$ from the above values of $\beta$, it can be shown that for a separating distance of $d=1 \mu m$ (where $C=8.85 \mu F/m^2$), the SPP field amplitude is identical to zero at the electrodes location $x=\pm \frac{d}{2}$ (i.e., $e^{-\alpha \frac{d}{2}}=e^{-34}$). We also consider the width $W=1 \mu m$, while the length $L$ is considered with different values.

\begin{figure}[ht!]\label{witnessvslength}
    \centering
    \subfloat[]{\includegraphics[width=5cm]{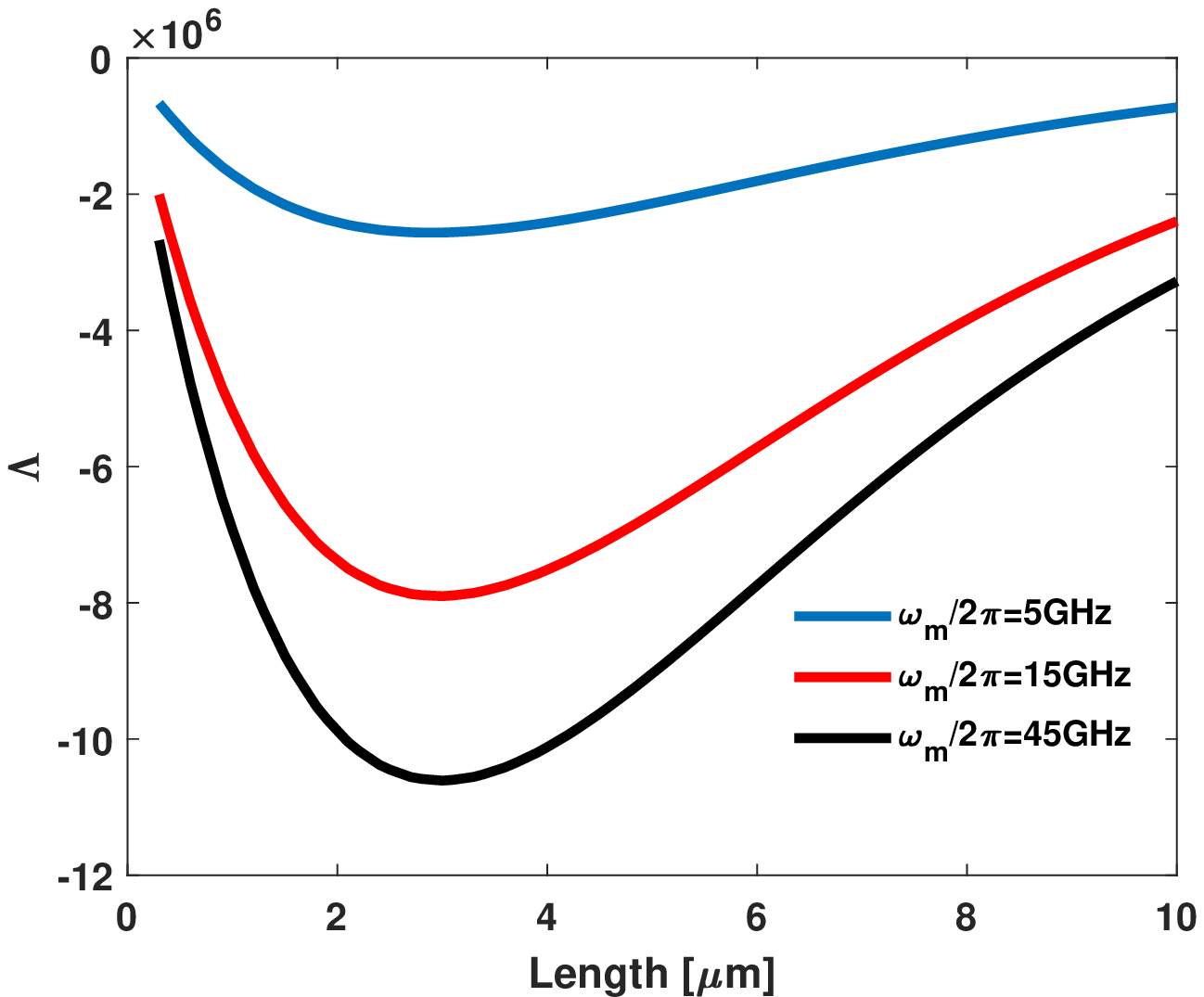}}%
    \quad
    \subfloat[]{\includegraphics[width=5cm]{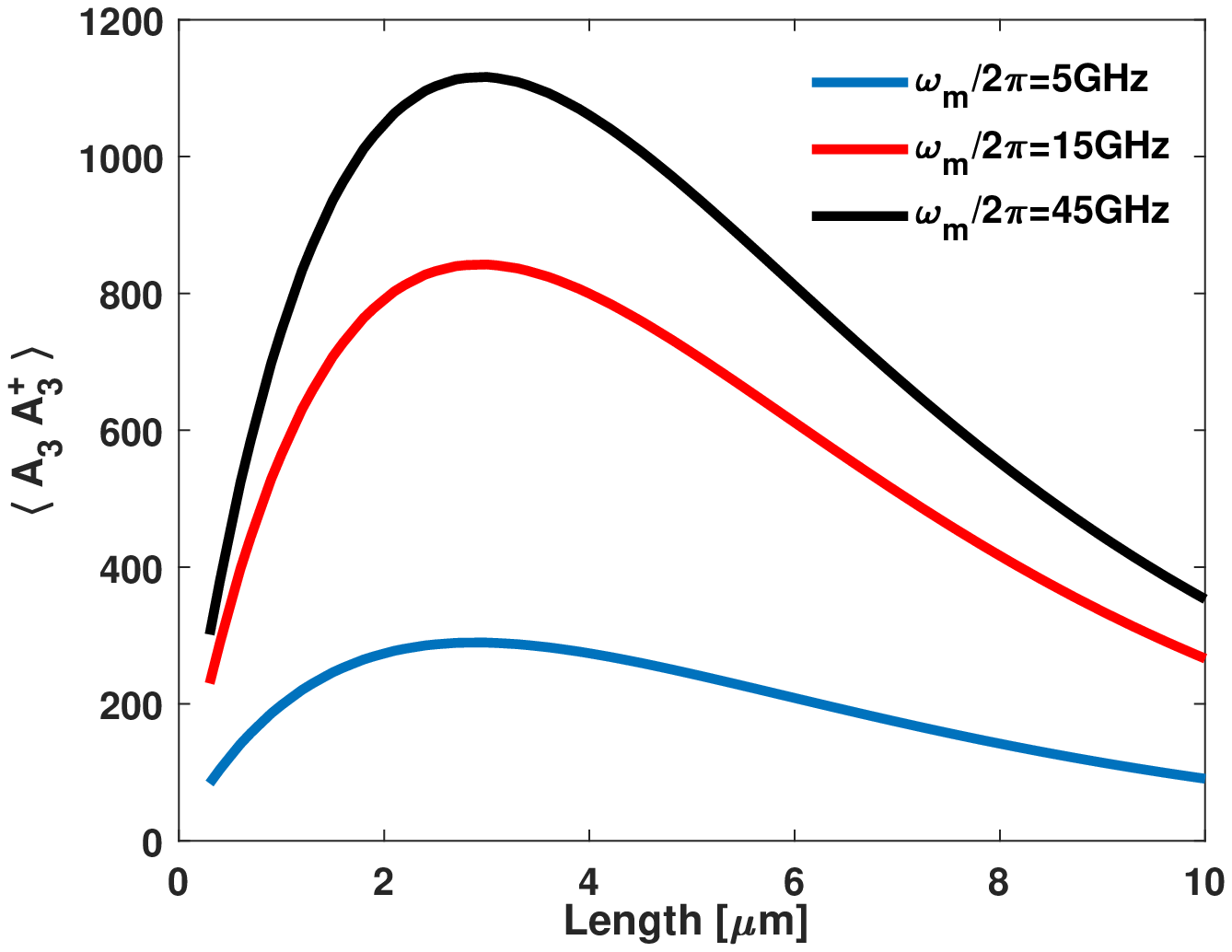}}%
    \caption{(a) The entanglement condition $\Lambda$ versus the interaction length. (b) The number of optical photons at $\omega_3$ versus the interaction length. Here $\lvert A_1\rvert\ ^2=10^6$.}
\end{figure}

In Fig. 3. (a), the entanglement condition $\Lambda$ is evaluated versus the waveguide length. Here, the optical pump intensity is $ \mid A_1  \mid^2=10^6$, the  microwave number of photons is $\hat{B}^{\dagger}\hat{B}|_{t=0}=10^4$, and three different microwave frequencies $\frac{\omega_m}{2\pi}=5 GHz$; $15 GHz$ and $45 GHz$ are considered. As can be seen, the fields are entangled for different waveguide lengths. However, the entanglement is stronger for larger microwave frequency. The entanglement strength is increasing against the waveguide length until losses start to take over. In Fig.3. (b), the number of generated photons at the lower sideband is calculated. We observe that significant number of photons are generated for optimum waveguide length. Limited by losses, both the entanglement and the number of generated photons at the lower sideband  have the same optimum waveguide length, $L=2.7 \mu m$.

\begin{figure}[ht!]\label{witnessvsA1A1}
    \centering
    \subfloat[]{\includegraphics[width=5cm]{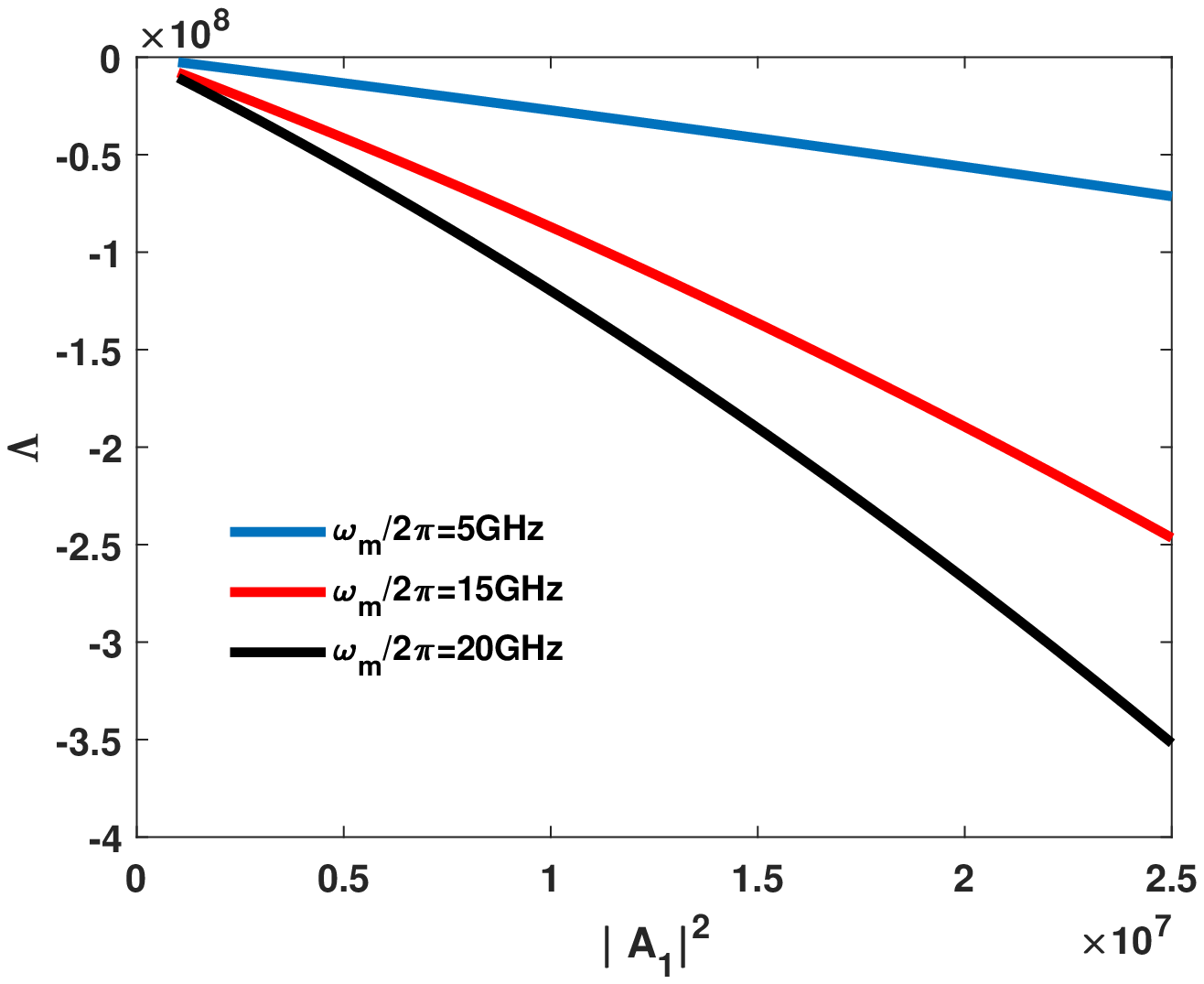}}%
    \quad
    \subfloat[]{\includegraphics[width=5cm]{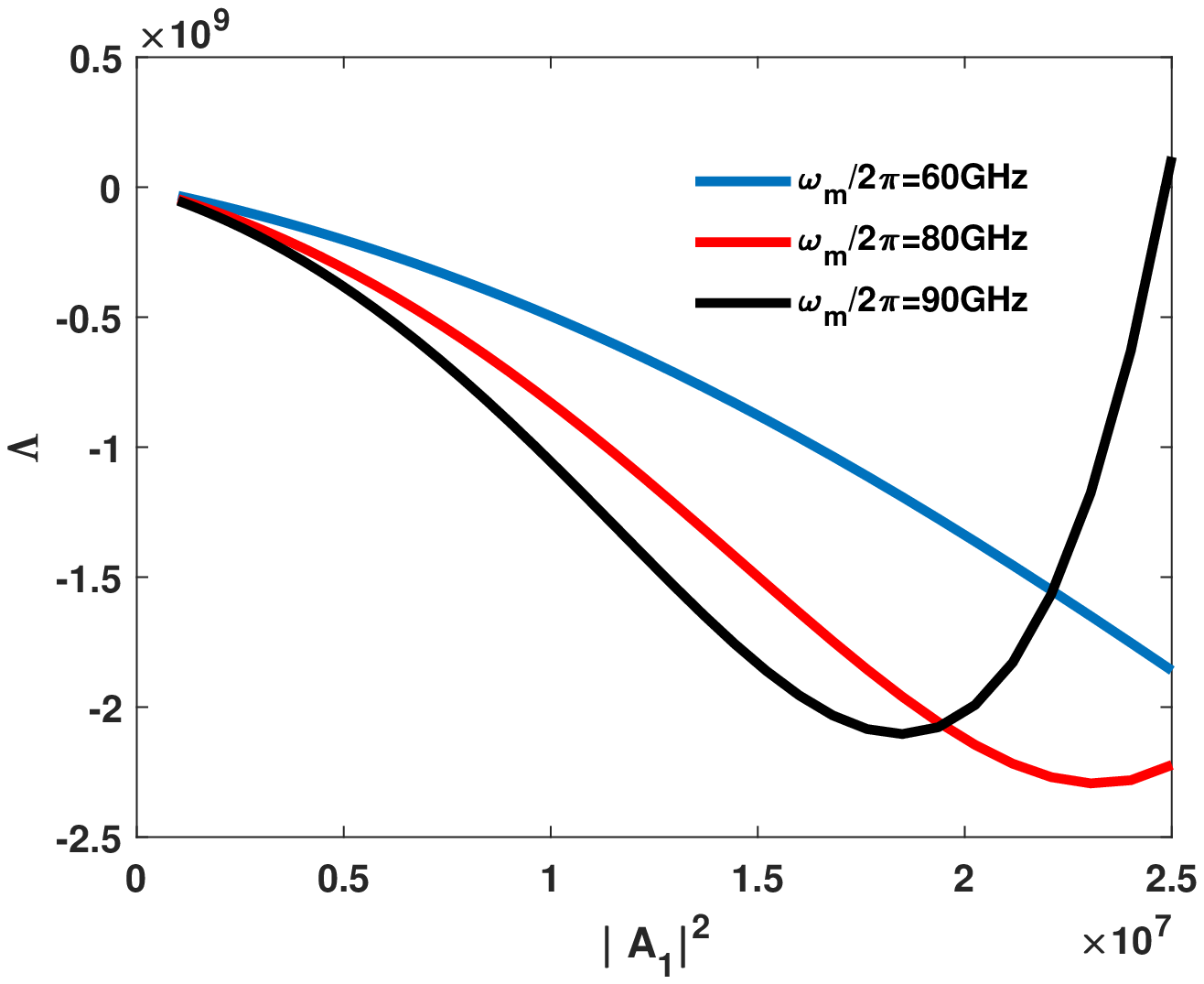}}%
    \caption{The entanglement condition $\Lambda$ versus the pump intensity $\lvert A_1\rvert\ ^2$. (a) The microwave frequencies are $\frac{\omega_m}{2\pi}$ =5GHz, 15 GHz and 20GHz. (b) The microwave frequencies are $\frac{\omega_m}{2\pi}$ =60GHz, 80 GHz and 90GHz. Here $L=2.7 \mu m $.}
\end{figure}

In Fig. 4, we have calculated the entanglement condition versus the optical pump intensity, considering the optimum waveguide length $L=2.7 \mu m$. Different microwave frequencies are considered. In Fig.4.(a), we consider $\frac{\omega_m}{2\pi}=$ 5 GHz ;15 GHz; and 20 GHz, while in Fig. 4 .(b) we consider $\frac{\omega_m}{2\pi}$=60 GHz; 80 GHz and 90 GHz. In both cases, the entanglement depends crucially on the pump intensity. For the microwave frequency values in Fig. 4 .(a),  the entanglement is stronger for larger pump intensities. However, for the higher microwave frequency values in Fig.4 .(b), the entanglement is maximized over specific pump intensity and gets weaker (up to vanishing) for larger intensities. For example, for $\frac{\omega_m}{2\pi}=$ 5 GHz, the entanglement is stronger for larger pump intensities over the considered range. While, for $\frac{\omega_m}{2\pi}=$90 GHz, the entanglement is maximal for $\lvert A_1\rvert\ ^2=1.8\times 10^7$, gets weaker for larger intensities, and the entanglement disappears for intensities greater than $ \lvert A_1\rvert\ ^2 =2.5\times 10^7$. 

\begin{figure}[ht!]\label{witnessvsBB}
    \centering
    \subfloat[]{\includegraphics[width=5cm]{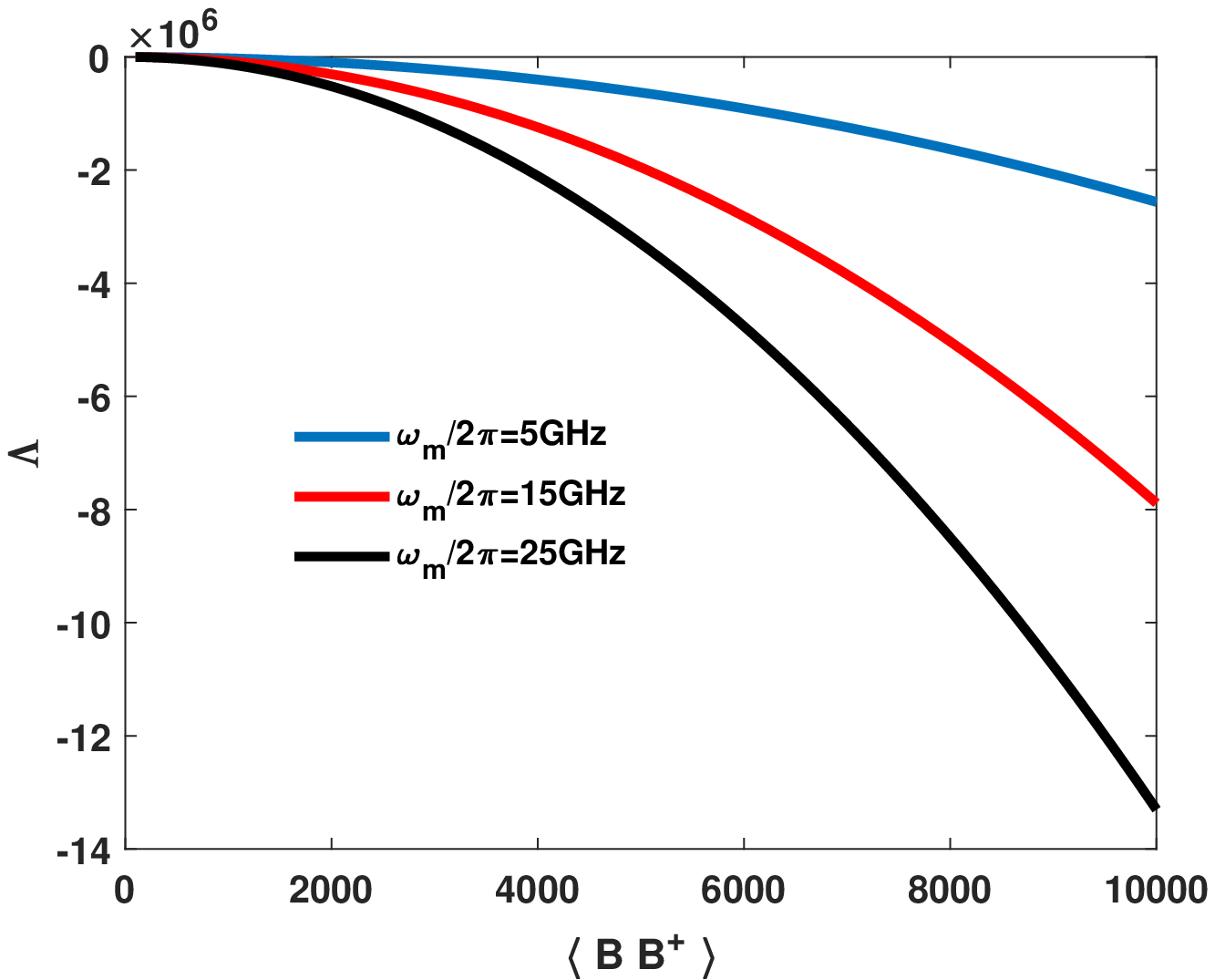}}%
    \quad
    \subfloat[]{\includegraphics[width=5cm]{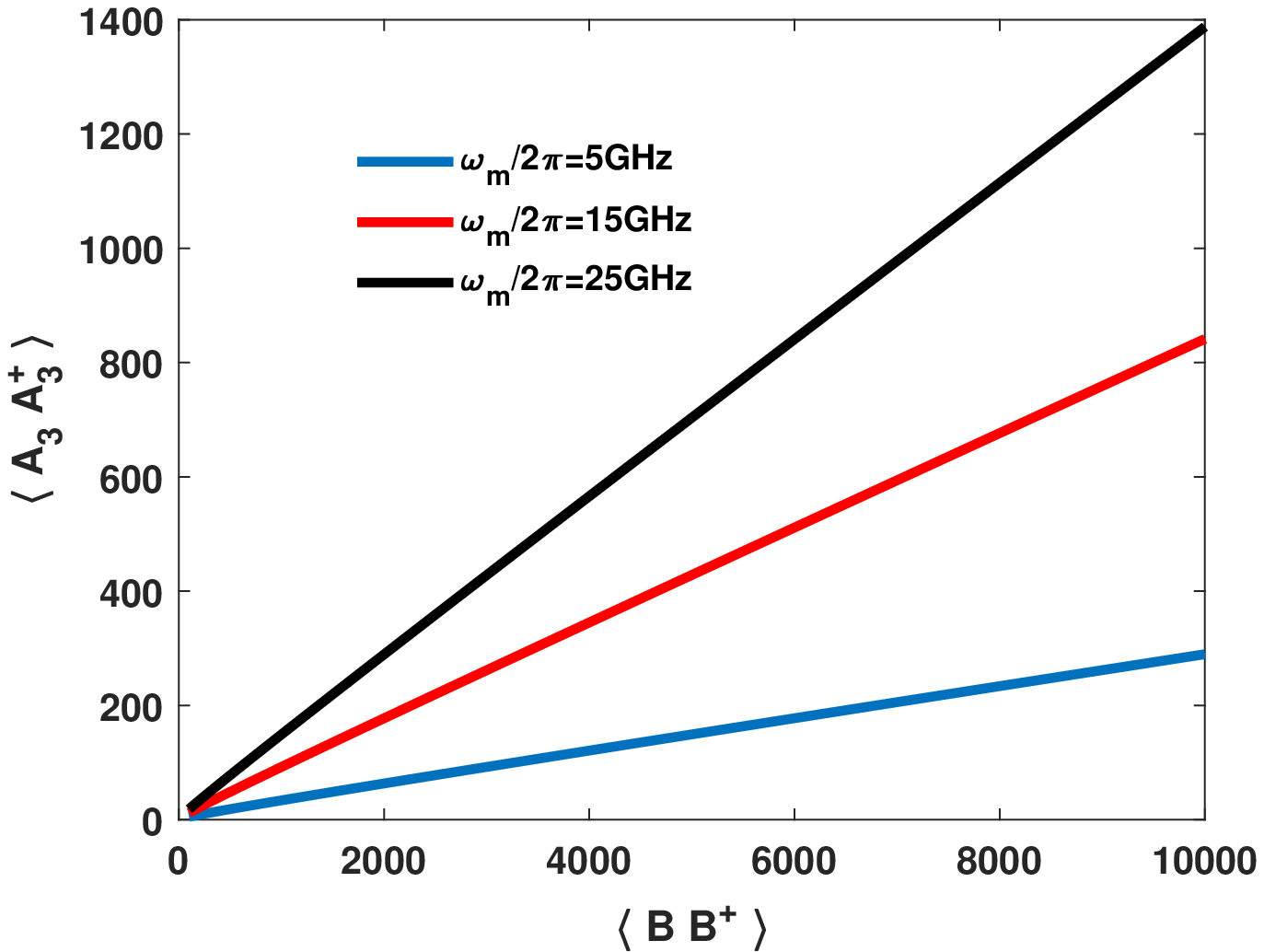}}%
         \caption{(a) The entanglement condition $\Lambda$ versus the microwave number of photons $\left\langle \hat{B}^{\dagger}\hat{B}\right\rangle$.  (b) The number of optical photons at frequency $\omega_3$ versus the microwave number of photons $\left\langle \hat{B}^{\dagger}\hat{B}\right\rangle$. Here  $\lvert A_1\rvert\ ^2=10^6$, and $L=2.7 \mu m$.}
\end{figure}

In Fig. 5, the entanglement condition, $\Lambda$, and the number of generated photon at the lower sideband are evaluated versus the microwave number of photons. We observe that the entanglement is stronger for larger number of microwave photons. This is also true for the number of photons generated at lower sideband. Different microwave frequencies are considered. Similar to the above observations, the entanglement strength  and number of generated photons get intensified for higher microwave frequency.           
\begin{figure}[ht!] \label{witnessvsfmfm}
    \centering
    \subfloat[]{\includegraphics[width=5cm]{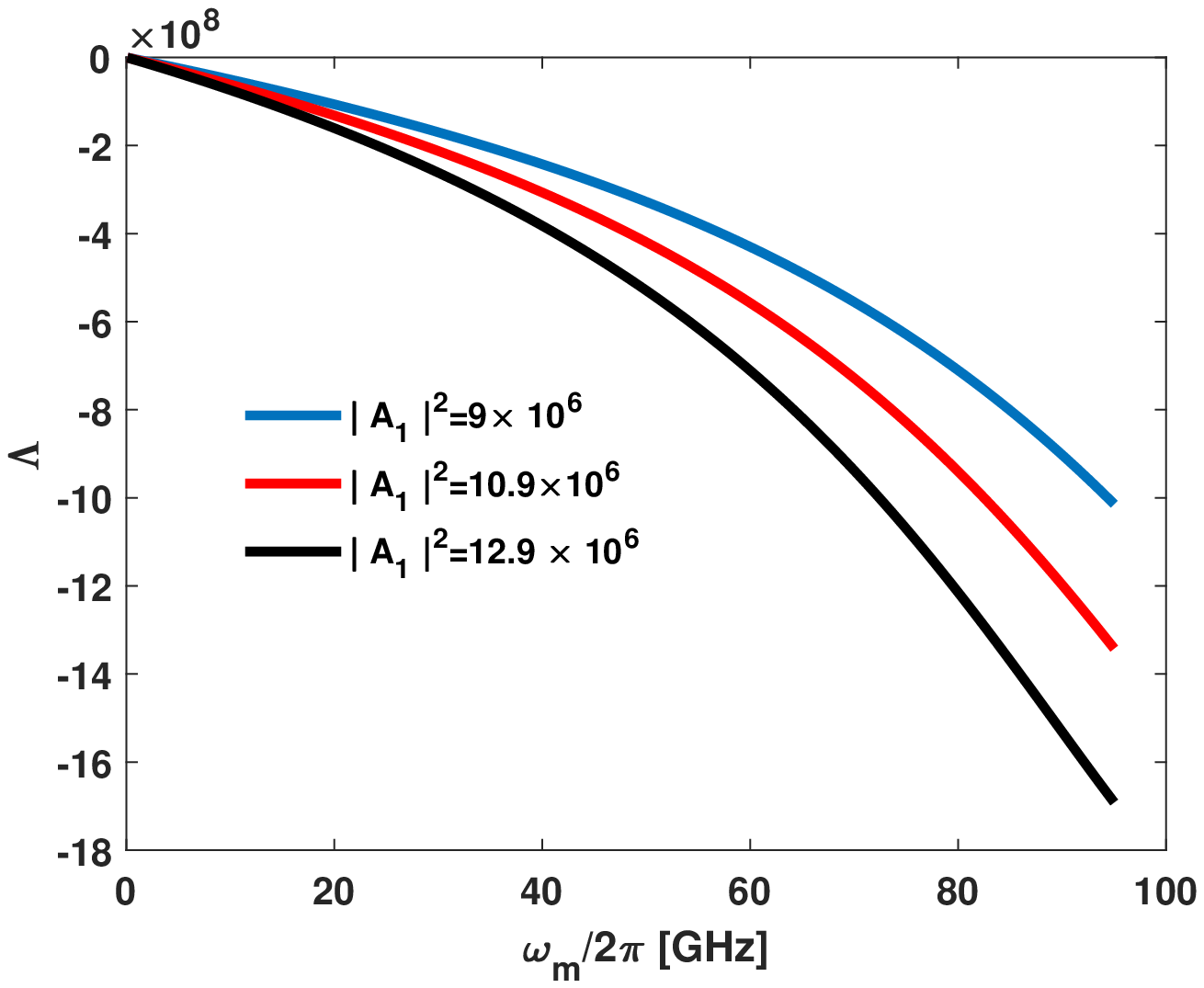}}%
    \quad
    \subfloat[]{\includegraphics[width=5cm]{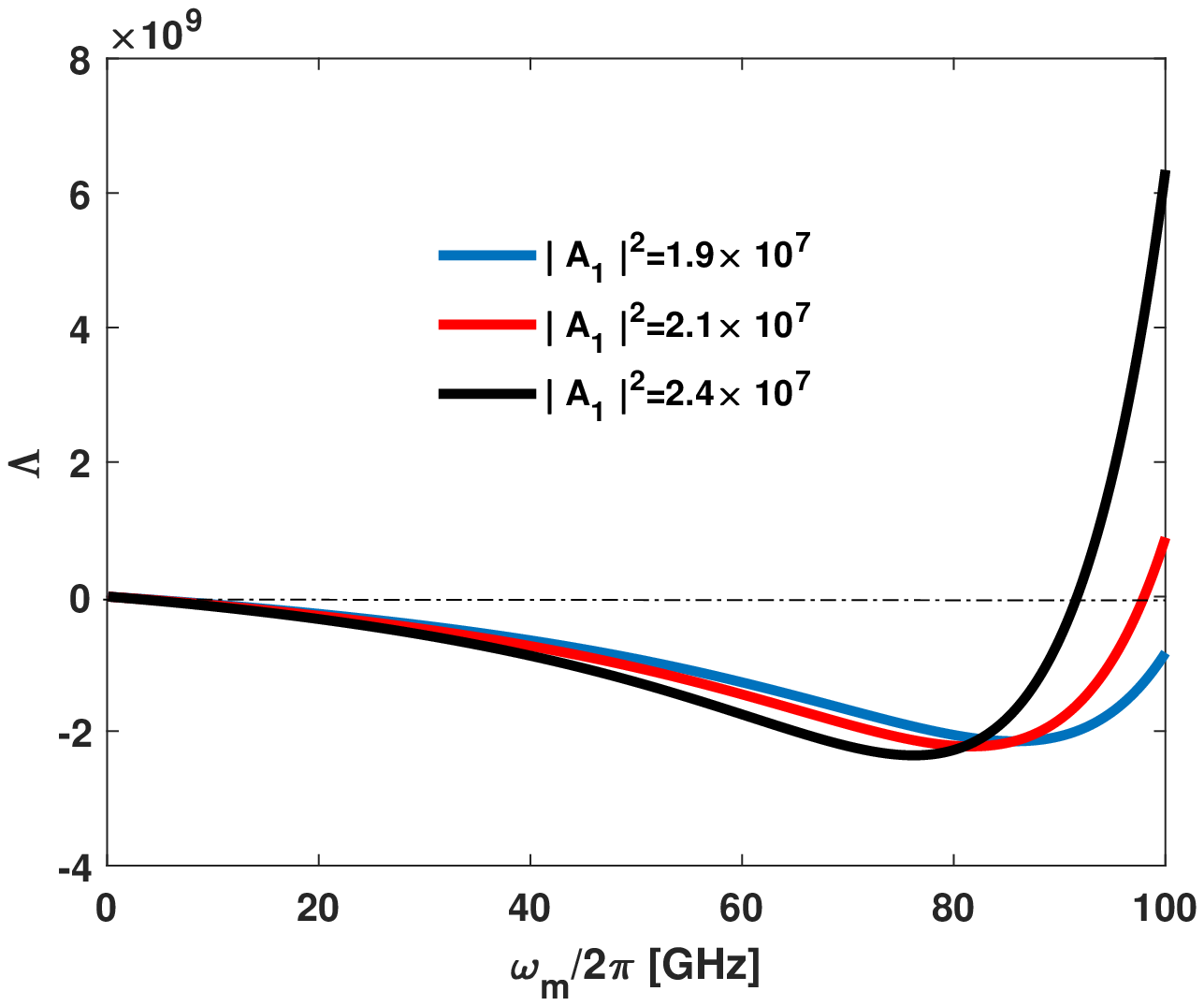}}%
     \caption{The entanglement condition $\Lambda$ versus the microwave frequency $\omega_m$. (a) The pump intensities are $\lvert A_1\rvert\ ^2$=$9\times 10^6$;$10.9\times 10^6$ and $12.9\times 10^6$. (b) The pump intensities are $\lvert A_1\rvert\ ^2$=$1.9\times 10^7$;$2.1\times 10^7$ and $2.4\times 10^7$.}
\end{figure}

In Fig.6., the entanglement condition $\Lambda$ is evaluated against the microwave frequency. Different pump intensities are considered. In Fig.6.(a), the entanglement is evaluated considering the intensities $\lvert A_1\rvert\ ^2=9\times 10^6$; $10.9\times 10^6$ and $12.9\times 10^6$. Also, in Fig.6 .(b), the pump intensities are $\lvert A_1\rvert\ ^2=1.9\times 10^7$; $2.1\times 10^7$ and $2.4\times 10^7$. For the pump intensity range in Fig.6. (a), the entanglement is stronger for higher microwave frequency and larger pump intensity. However, for the intensity range in Fig. 6. (b), the entanglement strength increases against the microwave frequency until reaching an optimum value and then starts to decrease until reaching no entanglement at specific microwave frequency. Both the optimum frequency and the frequency at which disentanglement is reached are smaller for larger pump intensity. However, using a larger pump intensity includes stronger entanglement. For example, for $\lvert A_1\rvert\ ^2=1.9\times 10^7$, the entanglement strength is maximal at the optimum microwave frequency $\frac{\omega_m}{2\pi}= 86$ GHz, and disentanglement is reached at $\frac{\omega_m}{2\pi}= 100$ GHz. However, for  $2.4\times 10^7$, the entanglement optimum frequency is $\frac{\omega_m}{2\pi}= 76$  GHz, and disentanglement is reached at $\frac{\omega_m}{2\pi}= 92$ GHz. Nonetheless, the entanglement at $\frac{\omega_m}{2\pi}= 76$ for  $\lvert A_1\rvert\ ^2=2.4\times 10^7$ is stronger than that at $\frac{\omega_m}{2\pi}= 86$ for  $\lvert A_1\rvert\ ^2=1.9\times 10^7$.

\section{Conclusion}

A microwave and optical fields entanglement based on electrical capacitor loaded with graphene plasmonic waveguide has been proposed and investigated. The microwave voltage is applied to the capacitor while the graphene waveguide is subjected to an optical surface plasmon polariton (i.e., SPP) input. It then follows that an SPP sidebands are generated at the expense  of the input SPP pump and the driving microwave signal. We have developed a quantum mechanics model to describe the fields interaction. The derived motion equations indicates entanglement between the microwave and the lower SPP sideband. Thus, we have applied the Duan's criterion to investigate the entanglement. The required equations needed to evaluate the Duan's determinant was derived from the motion equation using the quantum regression theorem. We found that the microwave signal and the lower SPP sideband are entangled over a vast microwave frequency. First, the entanglement was evaluate against the waveguide length. Limited by losses, it was observed that there is an optimum waveguide length at which the entanglement strength (and number of photons at the lower side band) are maximized. Second, we evaluated the entanglement versus the SPP pump intensity considering the obtained optimum length. It is found that the entanglement is stronger for larger pump intensity. However, for intensive pump inputs and microwave frequencies greater than 50 GHz, there is an optimum pump intensity at which the entanglement is maximized and then it decreases for larger intensity values until disentanglement is observed. Third, the entanglement is evaluated versus the microwave number of photons. As expected, the larger the number of microwave photons, the stronger the entanglement. Fourth, the entanglement was evaluated versus the microwave frequency. It is found that the entanglement is attained over the entire considered range. However, proper pump intensity must be provided. The proposed microwave-optical entanglement scheme is simple, compatible with the superconductivity and photonic technology, besides the major advantage of affording a frequency-tunable operation. 

\section*{Appendix A}

The chemical potential of the electrically driven graphene is given by $\mu_c=\hbar V_f\sqrt{\pi n_0+\frac{2C}{q} V_m}$. On following the same perturbation approach detailed in our previous work\cite{qasymehhichem} and  \cite{qasymehphase}, and by considering $C \mathcal{V} \ll \pi n_0 q $, the chemical potential can be approximated by:

\begin{equation} \label{eqB3}
	\mu_c=	\mu_{c}^{\prime}+ \mathcal{V} \mu_{c}^{\prime\prime} e^{-i 2 \pi f_m t}+c.c.,
\end{equation}
where $\mu_{c}^{\prime}=\hbar V_f\sqrt{\pi n_0}$, $\mu_{c}^{\prime\prime}=\hbar V_f\frac{C}{q \sqrt{\pi n_0}}$, $q$ is the electron charge, $n_0$ is the electron density per unit area, and $V_f=10^6$ m/s is the Fermi velocity of the Dirac fermions.

By substituting the chemical potential in Eq. (\ref{eqB3}) into the graphene conductivity expression $\sigma_{s}=\frac{iq^2}{4\pi\hbar}ln\bigg( \frac{2\mu_{c}-(\frac{\omega}{2\pi}+i\tau^{-1})\hbar}{2\mu_{c}+(\frac{\omega}{2\pi}+i\tau^{-1})\hbar}\bigg)+\frac{iq^2 K_B T}{\pi \hbar^2(\frac{\omega}{2\pi}+i\tau^{-1})}\bigg(\frac{\mu_{c}}{K_B T}+2 ln  \big( e^{-\frac{\mu_{c}}{K_B T}}+1\big) \bigg)$, and for $ \mathcal{V} \mu_{c}^{\prime\prime}\ll \mu_{c}^{\prime}$, the graphene's conductivity can be approximated up to the first order  \cite{qasymehphase} [29], yielding:    
\begin{equation} \label{eqB5}
	\sigma_s=\sigma_{s}^{\prime}+\mathcal{V} \sigma_{s}^{\prime\prime} e^{-i2 \pi f_m t}+c.c.,
\end{equation}

\begin{equation} \label{eqB6}
	\sigma_{s}^{\prime}=\frac{iq^2}{4\pi\hbar}ln\bigg( \frac{2\mu_{c}^{\prime}-(\frac{\omega}{2\pi}+i\tau^{-1})\hbar}{2\mu_{c}^{\prime}+(\frac{\omega}{2\pi}+i\tau^{-1})\hbar}\bigg)+\frac{iq^2 K_B T}{\pi \hbar^2(\frac{\omega}{2\pi}+i\tau^{-1})}\bigg(\frac{\mu_{c}^{\prime}}{K_B T}+2 ln  \big( e^{-\frac{\mu_{c}^{\prime}}{K_B T}}+1\big) \bigg),
\end{equation}

\begin{equation} \label{eqB7}
	\sigma_{s}^{\prime\prime}=\frac{iq^2}{\pi\hbar}\frac{(\frac{\omega}{2\pi}+i\tau^{-1})\hbar}{4(\mu_{c}^{\prime})^2-(\frac{\omega}{2\pi}+i\tau^{-1})^2\hbar^2}\mu_{c}^{\prime\prime}+\frac{iq^2 K_B T}{\pi \hbar^2(\frac{\omega}{2\pi}+i\tau^{-1})} tanh\bigg(\frac{\mu_{c}^{\prime}}{2K_B T}\bigg)  \frac{\mu_{c}^{\prime\prime}}{K_B T}.
\end{equation} where, $ \mathcal{V} \sigma_{s}^{\prime\prime}\ll \sigma_{s}^{\prime}$, $\hbar$ is the plank's constant, $\tau$ expresses the scattering relaxation time, $K_B$ represents the Boltzman constant, $T$ is the temperature, and $\omega$ is the frequency.

The classical Hamiltonian is given by:

\begin{equation} \label{eqB8}
\mathcal{H} = \frac{1}{2} \mathcal{V}^2 C \mathcal{A}_r+\frac{1}{2} \iiint_{V_L}\bigg( \varepsilon_{0}\varepsilon_{eff}\lvert \vec{E}_t\rvert\ ^{2} +\mu_0\lvert \vec{H}_t\rvert\ ^{2} \bigg) \partial V_L,
\end{equation}
where $\vec{E}_t=\sum_{j=1}^{3} \vec{E_j}$ and $\vec{H}_t=\sum_{j=1}^{3} \vec{H_j}$ are the total electric and magnetic fields associated with the SPP modes, $C=\frac{\varepsilon_{0}\varepsilon}{d}$ is the capacitance per unit area. 

On using the fields expressions in Eq. (\ref{eq1}), the effective permittivity expression in Eq. (\ref{eq4}), into the classical Hamiltonian in Eq. (\ref{eqB8}), one gets:

\begin{equation} \label{eqB9}
	\mathcal{H} = \mathcal{H}_0+\mathcal{H}_1,
\end{equation}
where $\mathcal{H}_0$ and $\mathcal{H}_1$ represent the classical free fields and interaction Hamiltonians, respectively, given by:

\begin{equation} \label{eqB10}
	\mathcal{H}_0 = \frac{1}{2}\mathcal{V}^2 C \mathcal{A}_r+\frac{1}{2} \mathcal{A}_r \sum_{j=1}^{3}  \lvert \mathcal{U}_j\rvert\ ^2 \bigg( \varepsilon_0 \varepsilon_{eff_j}^{\prime} \int_{\mathcal{}}^{}  \lvert \mathcal{D}_{x_j} \rvert\ ^2 \partial x+  \mu_0 \int_{\mathcal{}}^{}  \lvert \mathcal{D}_{y_j} \rvert\ ^2 \partial x \bigg)  ,
\end{equation}

\begin{equation} \label{eqB11}
\begin{split}
	\mathcal{H}_1 = &\frac{1}{2}\varepsilon_0 \varepsilon_{eff_2}^{\prime\prime}  \mathcal{U}_2^* \mathcal{V} \mathcal{U}_1 \Bigg[ \int_{\mathcal{-\infty}}^{+\infty} \bigg(  \mathcal{D}_{x_1} \mathcal{D}_{x_2}^*+\mathcal{D}_{z_1} \mathcal{D}_{z_2}^* \bigg)  \partial x \Bigg]\mathcal{A}_r Sinc\bigg( \frac{\beta_1-\beta_2}{2} L\bigg) e^{i(\beta_1-\beta_2)L} \\ +& \frac{1}{2}\varepsilon_0 \varepsilon_{eff_3}^{\prime\prime}   \mathcal{U}_1^* \mathcal{V} \mathcal{U}_3 \Bigg[\int_{\mathcal{-\infty}}^{+\infty} \bigg(  \mathcal{D}_{x_1}^* \mathcal{D}_{x_3}+ \mathcal{D}_{z_1}^* \mathcal{D}_{z_3} \partial x \bigg)\Bigg] \mathcal{A}_r Sinc\bigg( \frac{\beta_3-\beta_1}{2} L\bigg) e^{i(\beta_3-\beta_1)L}.
\end{split}
\end{equation}
Here, the SPP fields are considered independent of $y$, thus the result of integration with respect to $y$ is $W$, and $\int_{{0}}^{L} e^{i\Delta \beta z} \partial z = L Sinc\big( \frac{\Delta \beta L}{2}\big) e^{i\frac{\Delta\beta L}{2}}$ is used.

\section*{Appendix B}
By using the quantum regression theorem for Eqs.(\ref{eq13}-\ref{eq15}) multiple times, one can obtain the following closed set equations:
\begin{equation} \label{eqC4}
	\frac{ \partial\left\langle \hat{A}_3 \hat{B}\right\rangle}{\partial t}=-\frac{\Gamma_m}{2} \left\langle \hat{A}_3 \hat{B}\right\rangle- g_2 A^* \left\langle \hat{A}_3\hat{A}_2\right\rangle + g_3 A \left\langle\hat{A}_3 \hat{A}_3^\dagger \right\rangle,
\end{equation}

\begin{equation} \label{eqC5}
	\frac{ \partial\left\langle \hat{A}_3 \hat{A}_2\right\rangle}{\partial t}=-\frac{\Gamma_2}{2} \left\langle \hat{A}_3 \hat{A}_2\right\rangle+ g_2 A \left\langle \hat{A}_3\hat{B}\right\rangle,
\end{equation}

\begin{equation} \label{eqC6}
	\frac{ \partial\left\langle \hat{A}_3^\dagger \hat{A}_3\right\rangle}{\partial t}=-\frac{\Gamma_3}{2} \left\langle \hat{A}_3^\dagger \hat{A}_3\right\rangle+ g_3 A \left\langle \hat{A}_3^\dagger\hat{B}^\dagger\right\rangle,
\end{equation}

\begin{equation} \label{eqC7}
	\frac{ \partial\left\langle \hat{A}_3^\dagger \hat{B}^\dagger\right\rangle}{\partial t}=-\frac{\Gamma_m}{2} \left\langle \hat{A}_3^\dagger \hat{B}^\dagger\right\rangle- g_2 A \left\langle \hat{A}_3^\dagger\hat{A}_2^\dagger\right\rangle +g_3 A^* \left\langle \hat{A}_3^\dagger\hat{A}_3\right\rangle,
\end{equation}

\begin{equation} \label{eqC8}
	\frac{ \partial\left\langle \hat{A}_3^\dagger \hat{A}_2^\dagger\right\rangle}{\partial t}=-\frac{\Gamma_2}{2} \left\langle \hat{A}_3^\dagger \hat{A}_2^\dagger\right\rangle+ g_2 A^* \left\langle \hat{A}_3^\dagger\hat{B}^\dagger\right\rangle,
\end{equation}

\begin{equation} \label{eqC9}
	\frac{ \partial\left\langle \hat{B}^\dagger \hat{B}\right\rangle}{\partial t}=-\frac{\Gamma_m}{2} \left\langle \hat{B}^\dagger \hat{B}\right\rangle-g_2 A_1^* \left\langle \hat{B}^\dagger\hat{A}_2\right\rangle+ g_3 A_1 \left\langle \hat{B}^\dagger\hat{A}_3^\dagger\right\rangle,
\end{equation}

\begin{equation} \label{eqC10}
	\frac{ \partial\left\langle \hat{B}^\dagger \hat{A}_2\right\rangle}{\partial t}=-\frac{\Gamma_2}{2} \left\langle \hat{B}^\dagger \hat{A}_2\right\rangle+ g_2 A \left\langle \hat{B}^\dagger\hat{B}\right\rangle,
\end{equation}

\begin{equation} \label{eqC11}
	\frac{ \partial\left\langle \hat{B}^\dagger \hat{A}_3^\dagger\right\rangle}{\partial t}=-\frac{\Gamma_m}{2} \left\langle \hat{B}^\dagger \hat{A}_3^\dagger\right\rangle+ g_3 A^* \left\langle \hat{B}^\dagger\hat{B}\right\rangle,
\end{equation}

These equations can be solved using an iterative approach for a given initial conditions.   

\noindent\textbf{Disclosures.} The authors declare no conflicts of interest.

\end{document}